\documentclass[intlimits,twoside,a4paper]{article}

\usepackage{amsmath,amssymb}
\usepackage{graphicx}
\usepackage[T2A]{fontenc}
\usepackage[cp1251]{inputenc}

\usepackage{cmpj2}

\issue{2014}{17}{2}{23603}
\doinumber{10.5488/CMP.17.23603}

\title[Solvent primitive  double layer model]
{Solvent primitive model of an electric double layer in slit-like pores: microscopic structure,
adsorption and capacitance from a density functional approach}

\author[O. Pizio, S. Soko{\l}owski]{O. Pizio\refaddr{label1}\thanks{E-mail: oapizio@gmail.com}\ , S. Soko{\l}owski\refaddr{label2}}
\addresses{
\addr{label1} Instituto de Qu\'{i}mica, Universidad Nacional Autonoma de M\'{e}xico,
Circuito Exterior, Ciudad Universitaria, \\ 04510 M\'{e}xico, D.F., M\'{e}xico
\addr{label2} Department for the Modelling of Physico-Chemical Processes,
Maria Curie-Sklodowska University, \\Gliniana~33, Lublin, Poland}

\date{Received January 27, 2014, in final form February 23, 2014}
\authorcopyright{O. Pizio, S. Soko{\l}owski, 2014}

\begin{document}
\maketitle

\begin{abstract}
We investigate the electric double layer formed between charged walls of a slit-like pore
and a solvent primitive model (SPM) for electrolyte solution. The recently developed
version of the weighted
density functional approach for electrostatic interparticle interaction is applied to
the study of the density profiles, adsorption and selectivity of adsorption of ions and
solvent species. Our principal focus, however,
is in the dependence of differential
capacitance on the applied voltage, on the electrode and on the pore width.
We discuss
the properties of the model with respect to the behavior of a primitive model,
i.e., in the absence of a  hard-sphere solvent.
We observed that the differential capacitance of the SPM on the applied electrostatic potential
has the camel-like shape unless the ion fraction is high. Moreover, it is documented
that the dependence of differential capacitance of the SPM on the pore
width is oscillatory, which is in close similarity to the primitive model.

\pacs  61.20.Gy, 61.20.Qg, 65.20.Jk, 68.08.-p, 68.43.De
\end{abstract}

\section{Introduction}
The most frequently applied microscopic modelling for the electric double layer (EDL)
formed at an interface between a charged solid surface
involves the primitive model (PM) of the fluid ionic subsystem.
Namely, it is assumed that ions are charged hard spheres immersed into
a dielectric continuum having a certain dielectric constant. This
very simplified model, compared to real systems in laboratory,
has been used for the development and testing of
theoretical approaches, as well as
to explain experimental observations.

In the theoretical approaches, the dielectric discontinuity at
the electrode-electrolyte interface
is usually neglected. Another simplification commonly used in the problem of adsorption
of PM electrolyte solutions into slit-like pores is to assume that the dielectric
constant of the bulk fluid and inside a pore is the same. These comments just
illustrate how far the present theoretical modelling is from
an entirely satisfactory description of the EDL problems.

One step forward can be made by considering
the solvent primitive model (SPM) rather than the PM  in the EDL problems.
The essence of the SPM is to take into account the effects of excluded volume,
due to the presence of solvent molecules
(most frequently considered as
hard spheres) that are neglected
in the PM. First attempts to investigate the SPM at a charged surface have been undertaken
in references \cite{grimson,groot}. More comprehensive
efforts to explore the properties of the SPM
at charged surfaces have been carried out
using a density functional theory~\cite{tang1,hansen1} and
Monte Carlo computer simulations~\cite{boda1,stasio1}.
For the purposes of our study, it is worth mentioning that Tang et al.~\cite{tang1} used
Tarazona's weighted density method
to describe the hard sphere interaction, while
the electrostatic contribution to the free energy functional
was modelled assuming that
the residual part of the direct correlation functions of nonuniform fluid is
the same as in a bulk ionic system.
On the other hand, in their recent
investigation, Oleksy and Hansen~\cite{hansen1}
used a version of the density functional approach in which the electrostatic correlation
contribution was neglected.
In the same context, quite recently the SPM has been used
to describe certain aspects of partitioning of electrolyte solutions through semipermeable
membranes~\cite{boda2,boda3,henderson1}.
The importance of such sophistication of modelling, in spite of intrinsic impossibility
to describe dielectric properties of the solvent medium, has been documented.

The present state of knowledge regarding the properties of the SPM electric double layer is still
incomplete, in particular, concerning the problem of adsorption of electrolyte solutions in the slit-like
pores, where an overlap of structures formed at two pore walls can cause some peculiarities
of the density profiles, adsorption, dependence of the
accumulated charge on the applied voltage and differential capacitance.
The overlap of double layers formed at each wall has been involved in the interpretation of
recent experimental
observations of the dependence of the capacitance of an electrolyte solution on the pore
width~\cite{chmiola1,chmiola2}, exhibiting a maximum for a particular very narrow pore of the width
slightly larger than the value of the diameter of ions. Computer simulations performed for
primitive type models, though with sophistication of the internal structure of ions in some cases, have confirmed the
experimental results and provided a certain explanation of the peculiarities of the behavior of the
differential capacitance in narrow pores~\cite{merlet,georgi,kondrat,wu,feng}.

The study of the effect of the differential capacitance of the SPM electric double layer
on the value of electrostatic potential, on the pore walls and on the pore width
is the principal issue of the present communication. To investigate this model, we use
the recent successful weighted density functional approach proposed for a
restricted primitive model of electrolyte solutions in contact with charged solid
surface~\cite{wang1,wang2}. Here, this approach is extended to a mixture of positive and negative ions and
hard spheres confined in slit-like pores. In doing this, we use the recent developments
dealing with the study of a similar problem, although at the level of the PM for electrolyte
solutions in slit-like pores~\cite{pizio1,pizio2,jiang,doug}.

\section{The model and theory}

The SPM under consideration consists of three
species, i.e.,  positive and negative ions ($+,-$) and solvent molecules
mimicked as hard spheres (hs).
For the sake of simplicity, in this work
we assume that the diameters of all species
are the same,  $\sigma_+=\sigma_-=\sigma_\mathrm{hs}=\sigma$. The valencies of cations and anions
are the same $Z^{(+)}=|Z^{(-)}|=Z$. Moreover, we restrict to univalent ions in what follows,
i.e., $Z=1$.  The interactions between species are as follows:
\begin{equation}
u^{(\alpha\gamma)}(r)=\left\{
\begin{array}{ll}
\infty , & r<\sigma,\\
\frac{{e}^{2}Z^{(\alpha )}Z^{(\gamma) }}{4\pi \epsilon\epsilon_0}\frac{1}{r}, & r>\sigma,
\end{array}
\right.   \label{eq:33}
\end{equation}
where $\alpha$, $\gamma = +, -, \text{hs}$; $e$ denotes
the magnitude of elementary charge, $\epsilon$ is the relative permittivity and
 $\epsilon_0$  is the permittivity of the vacuum.
Also $Z^{(\textrm{hs})}=0$, thus the solvent is just the fluid of hard spheres.

The mixture of three components is confined in a slit-like pore of the width $H$.
The interaction of ions with the pore walls
is described by the potential $v^{(\alpha)}(z)=v'^{(\alpha)}(z)=v'^{(\alpha)}(H-z)$
($\alpha=+, -$),
\begin{equation}
v'^{(\alpha)}(z)=v_\mathrm{hw}(z) + v_\mathrm{el}^{(\alpha)}(z),
\label{eq:2}
\end{equation}
 where
$ v_\mathrm{hw}(z)$ is the hard-wall potential
\begin{equation}
 v_\mathrm{hw}(z)=\left\{
\begin{array}{ll}
 \infty, & \text{for $z < \sigma/2$ and  $z > H-\sigma/2$},  \\
0, & \text{otherwise},
\end{array}
\right.
\label{eq:2a}
\end{equation}
and
\begin{equation}
\beta v_\mathrm{el}^{(\alpha)}(z)=-2\pi l_\mathrm{B} Q Z^{(\alpha)}  z
\label{eq:2b}
\end{equation}
is the Coulomb potential. In the above $\beta=1/kT$,
$Qe$ is the surface charge density of the wall, $l_\mathrm{B}= e^2/(4\pi kT \epsilon\epsilon_0)$ denotes the
Bjerrum length. Energetic aspects of interactions between ions for the model in hand
are given in terms of reduced temperature $T^*_\mathrm{el}=\sigma/l_\mathrm{B}$.
We assume that the interaction of solvent species with the pore walls,
$v^{(\mathrm{hs})}(z)=v'^{(\mathrm{hs})}(z)=v'^{(\mathrm{hs})}(H-z)$,  is given in the form of Yukawa potential,
\begin{equation}
  v'^{(\mathrm{hs})}(z)=\left\{
\begin{array}{ll}
 \infty, & \text{for $z < \sigma/2$  and $z > H-\sigma/2$},  \\
\varepsilon_\mathrm{gs} \exp[-\lambda_\mathrm{gs}(z-\sigma/2)]/z, & \text{otherwise}.
\end{array}
\right.
\label{eq:2c}
\end{equation}

The confined mixture is in equilibrium with the bulk mixture composed of
the same components. The bulk dimensionless densities of
the species $\alpha=+, -, \mathrm{hs}$  are $\rho^*_{\alpha}=\rho_{\alpha}\sigma^3$
($\rho^*_\mathrm{ion}=\rho^*_+ +\rho^*_-$).

We use the density functional approach, described
more in detail in our recent works, see e.g., references~\cite{pizio1,pizio2,bryk1}.
In essence, we construct a
thermodynamic potential for the system and then
the equilibrium density profiles are obtained by
minimizing the thermodynamic potential,
\begin{equation}
 \Omega= F+
\sum_{\alpha=+,-,\mathrm{hs}}\int \rd\mathbf{r}  \left[v^{(\alpha)}(z)\rho^{(\alpha)}(z) -\mu_{\alpha}\right]
+\int \rd\mathbf{r}q(z)\Psi(z).
\end{equation}
In the above
$\rho^{(\alpha)}(z)$  and
$\mu_\alpha$ are the local density and the chemical potential of the species $\alpha$, respectively,
$F$ is the free energy functional and
$q(z)$ is the charge density,
\begin{equation}
 q(z)/e=\sum_{\alpha=+,-}Z^{(\alpha)}\rho^{(\alpha)}(z).
\end{equation}
The electrostatic $\Psi(z)$ satisfies the Poisson equation,
\begin{equation}
 \nabla^2\Psi(z)=-\frac{4\pi}{\epsilon\epsilon_0}q(z).
\label{eq:poi}
\end{equation}
The solution of differential equation (\ref{eq:poi})
for the slit-pore geometry with walls of equal charge is perfomed
similarly to reference \cite{henderson2}, where the model is different, however.
Moreover, the method of solving the Poisson equation for a set  of interconnected slit-like pores
with permeable walls was explained and analysed in every detail
in the recent work by Kovacs et al.~\cite{kovacs}.
For the model of a single slit defined by equations~(\ref{eq:2})--(\ref{eq:2c}) in the present study, the
solution requires the choice of the boundary condition, namely
of the value of the electrostatic potential at a wall, $V_0=\Psi(z=0)=\Psi(z=H)$.

From the electro-neutrality condition of the
system it follows that
\begin{equation}
 Q+\int \rd z q(z) = 0,
 \label{eq:neutral}
\end{equation}
where $Qe$ is the surface charge density of the wall as we have already mentioned above.

The  free energy of the system, $F$, is
the sum  of the ideal, $F_\mathrm{id}$, hard sphere, $F_\mathrm{hs}$
and residual electrostatic excess contribution, $F_\mathrm{el}$,
arising from the coupling between electrostatic and
hard-sphere interactions.
The ideal part of the free energy, $F_\mathrm{id}$, is known exactly,
\begin{equation}
 F_\mathrm{id}/kT=\sum_{\alpha=a,c,\mathrm{hs}}\int \rd{\bf r}\rho^{(\alpha)}(z)\left[\ln\left(\rho^{(\alpha)}(z)\right)-1\right].
\end{equation}
The excess free energy due to hard sphere interactions between the species $+,-$, and hs, $F_{\mathrm{hs}}$,
is taken from the fundamental measure theory \cite{yuwu,yuwu2,yuwu3,ros1,ros2}.
The details of the White Bear version of the fundamental measure theory are given in references~\cite{yuwu,yuwu2,yuwu3}
Finally, the residual electrostatic contribution $F_\mathrm{el}$ is described by using the so-called
``weighted correlation approach'', WCA-$k^2$ approximation, developed
for nonuniform RPM ionic fluids by Wang et al. \cite{wang1,wang2} based
on the analytic solution of the mean spherical approximation, cf. also reference~\cite{pizio1}.
The expressions used in the present study are given by equations~(11)--(14) of our recent work~\cite{pizio1}.
They are omitted to avoid the unnecessary repetition.

At equilibrium, the density profiles minimize the thermodynamic potential $\Omega$,
i.e.,
\begin{equation}
 \frac{\delta \Omega}{\delta\rho^{(\alpha)}(\mathbf{r})}=0, \qquad \alpha=+,-,\mathrm{hs}.
 \label{eq:minim}
\end{equation}
The resulting density profile equations can be straightforwardly
derived by modifying those given in references \cite{pizio1,wang1,wang2,hansen1}

As we have mentioned above, the adsorption system is in equilibrium
with the bulk SPM mixture, this equilibrium being determined by the equality
of chemical potentials of each species in the bulk phase and in the pore.
The bulk  densities of ionic species
satisfy the electro-neutrality condition $Z^{(+)}\rho_{+}+Z^{(-)}\rho_{-}=0$.

\section{Results}

Let us now specify a set of parameters of the model we study below.
As already mentioned in the introduction, we restrict our attention to the model of equal
diameters of all the species involved. Actually, we performed calculations for the
model with a larger diameter of solvent species compared to the diameter of ions, but
qualitatively the trends observed are very similar to those discussed below.
The distance from the wall and pore width are given in reduced units, $z^*=z/\sigma$ and
 $H^*=H/\sigma$, respectively. Also, the electrostatic potential at the wall is
considered in reduced units, $V^*=eV_0/kT$.
Another introductory comment concerns the interaction between solvent hard sphere species
and pore walls. It has been written in the form of Yukawa interaction. However, in the
present study we just consider weakly adsorbing walls, $kT/\varepsilon_\mathrm{gs}=1$ and
$\lambda_\mathrm{gs}=3$. This interaction has been introduced having in mind a possible extension
of the SPM model in order to take into account the attractive interaction between solute particles
in the spirit of works by Oleksy and Hansen~\cite{hansen1,hansen2,hansen3,hansen4}.
On the other hand, our interest is in a dense fluid with high fraction of solvent species
and low ion content. Thus, in the majority of numerical calculations,  the solvent bulk density
is taken to be $\rho^*_\mathrm{hs}=0.5$.

\subsection{Density profiles and adsorption}

\begin{figure}[!b]
\begin{center}
\includegraphics[width=0.44\textwidth,clip]{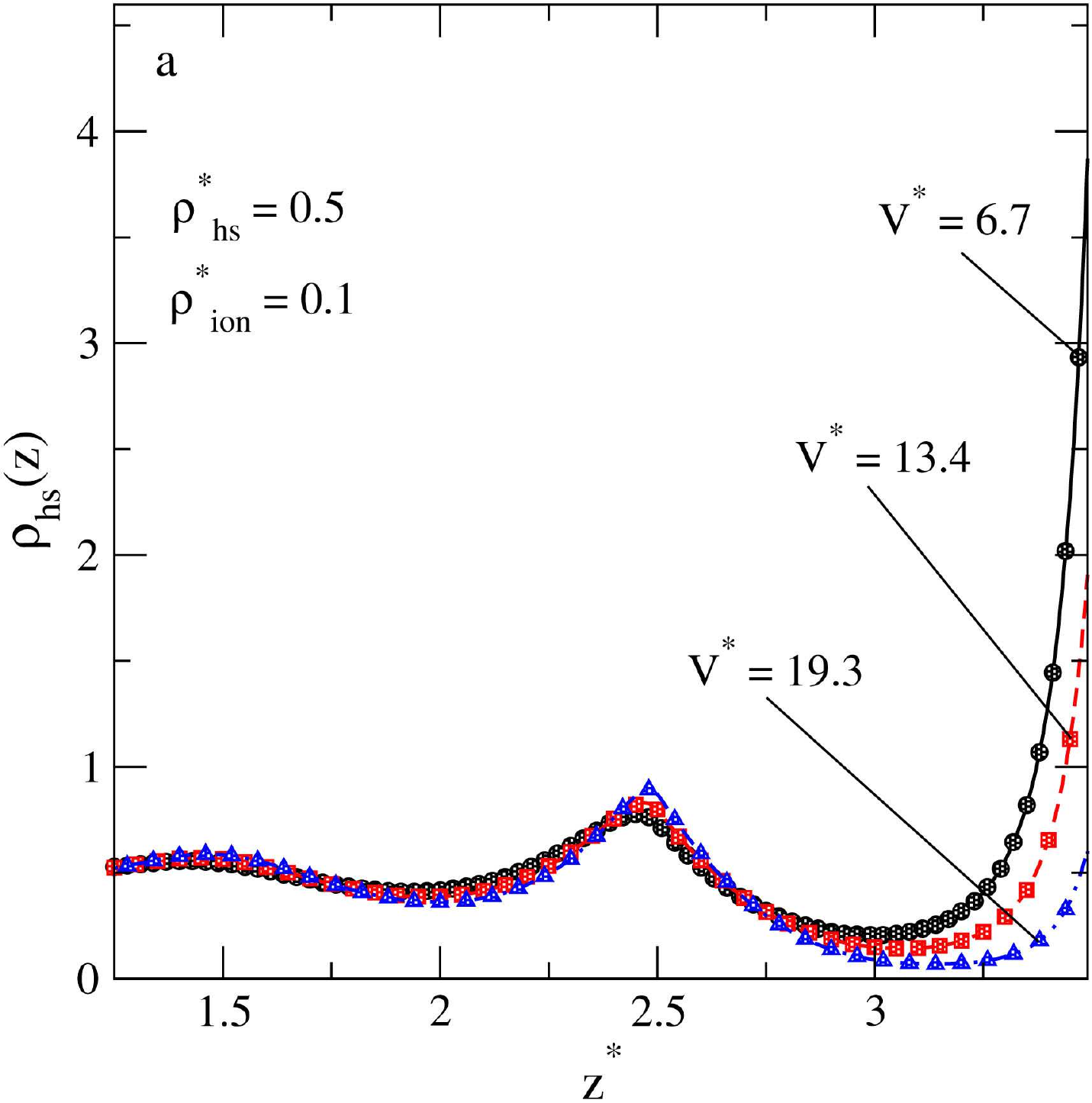}
\hspace{3mm}
\includegraphics[width=0.46\textwidth,clip]{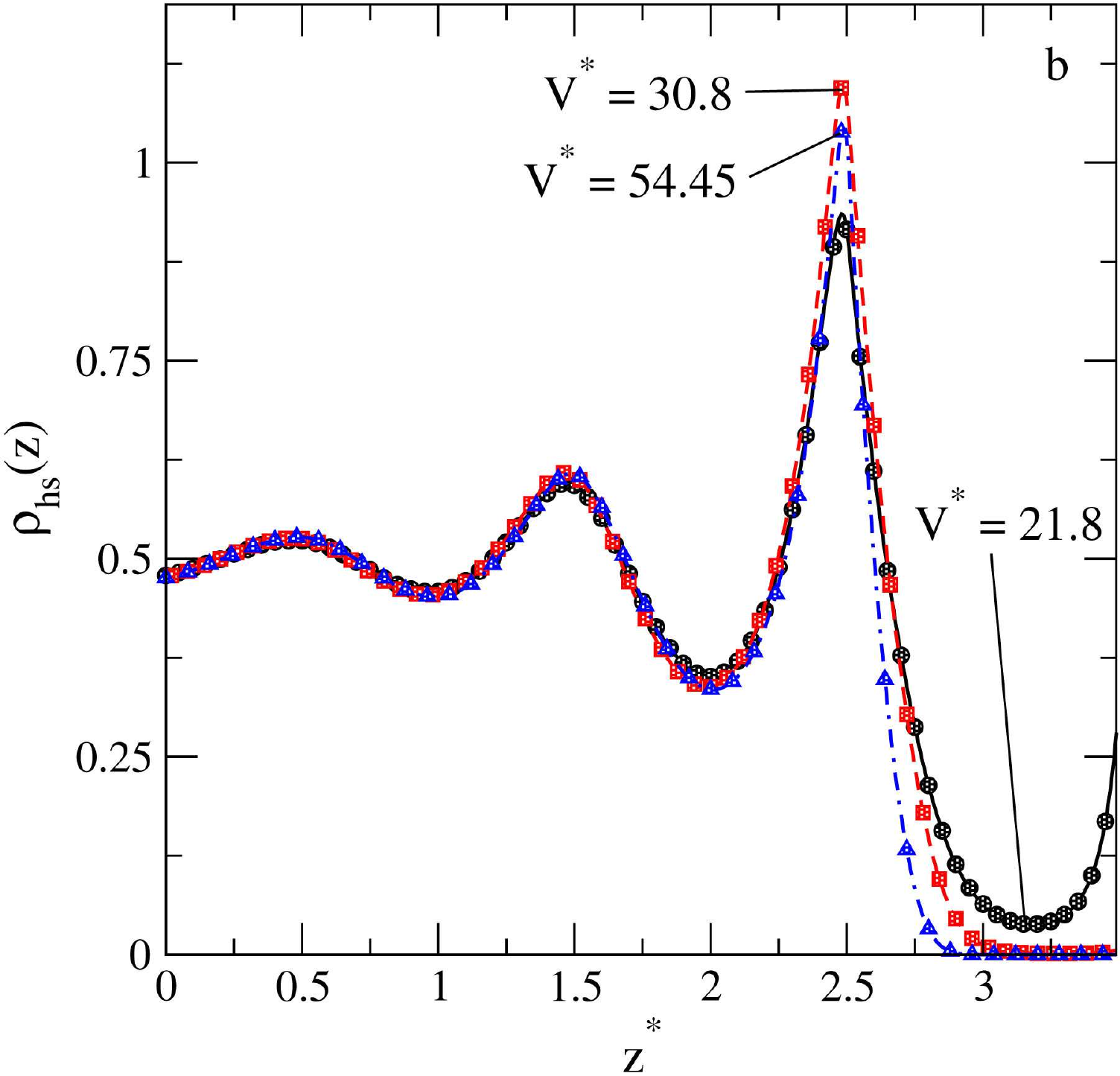}
\end{center}
\vspace{-2mm}
\caption{(Color online) Evolution of the density profiles of hard sphere species, $\rho_\mathrm{hs}(z)$,
of the SPM, with the applied electrostatic potential on the wall, $V^*$,
at bulk density, $\rho^*_\mathrm{hs}=0.5$, $\rho^*_\mathrm{ion}=0.1$,
in the slit-like pore of the width $H^*=8$. The energetic parameters of the SPM are
$T^*_\mathrm{el}=0.15$, $kT/\varepsilon_\mathrm{gs} = 1.0$ here and in all the subsequent figures.\label{fig1}}
\protect
\end{figure}

\begin{figure}[!t]
\begin{center}
\includegraphics[width=0.45\textwidth,clip]{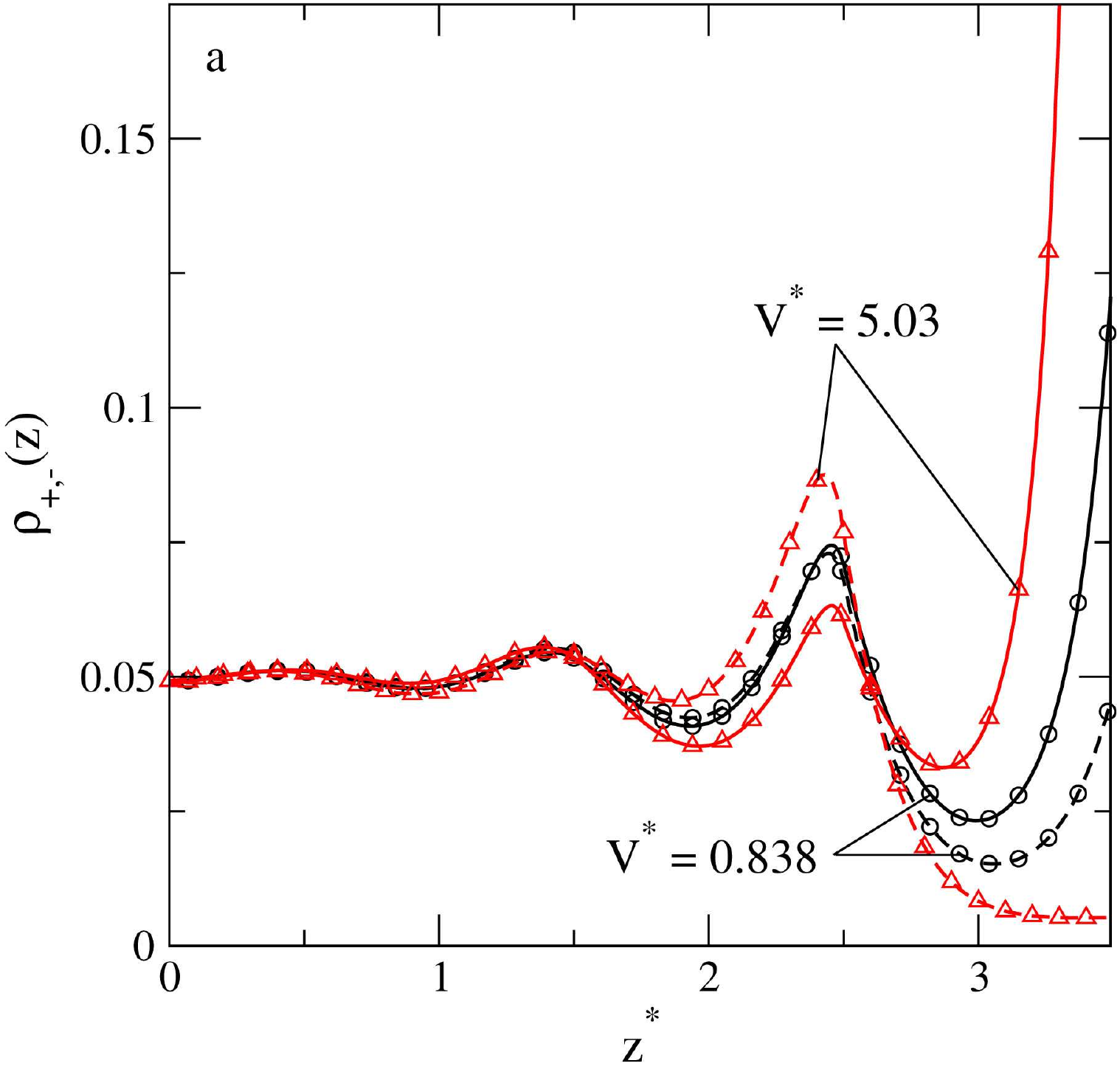}\hspace{3mm}
\includegraphics[width=0.45\textwidth,clip]{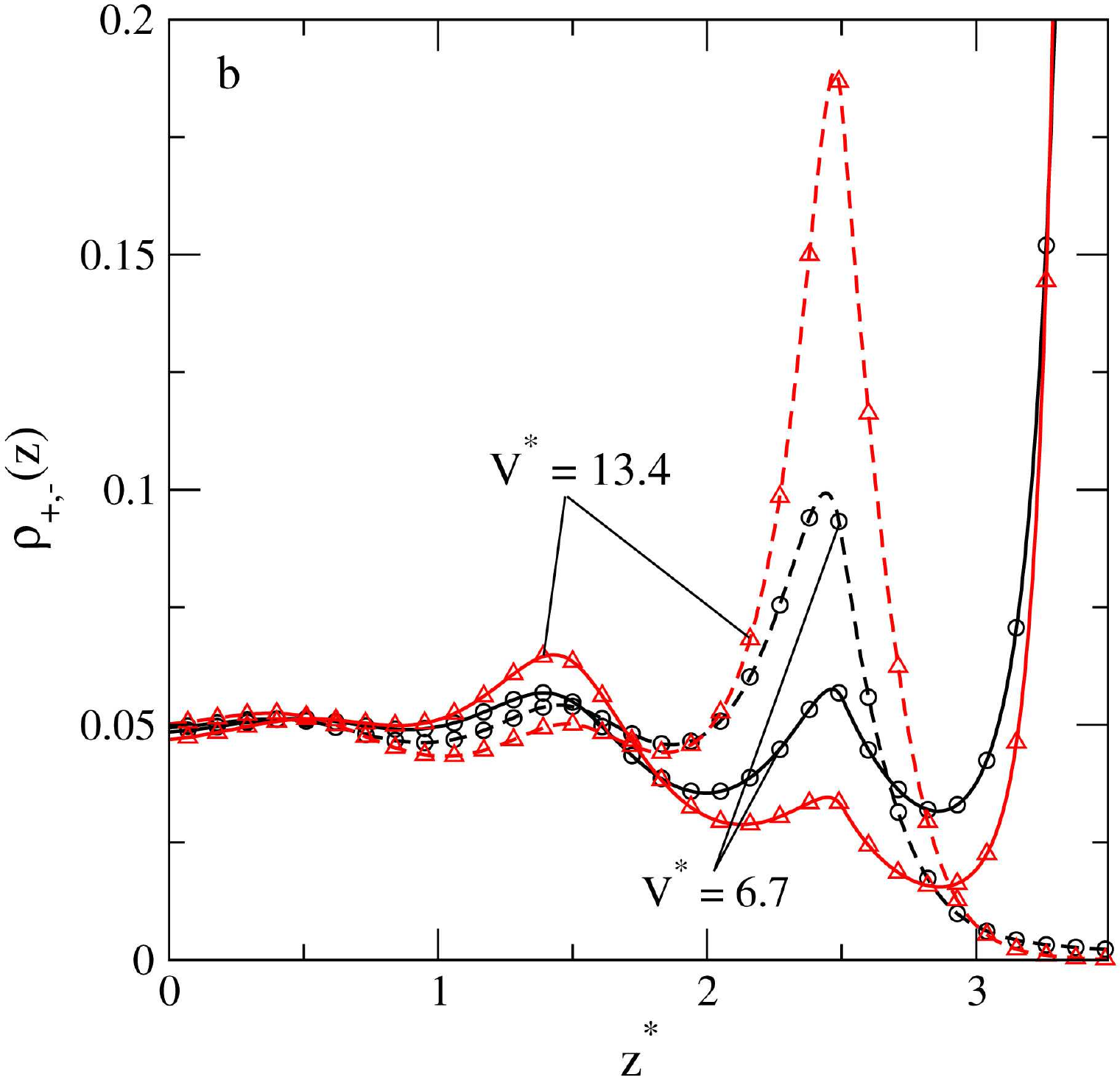}\\[1ex]
\includegraphics[width=0.45\textwidth,clip]{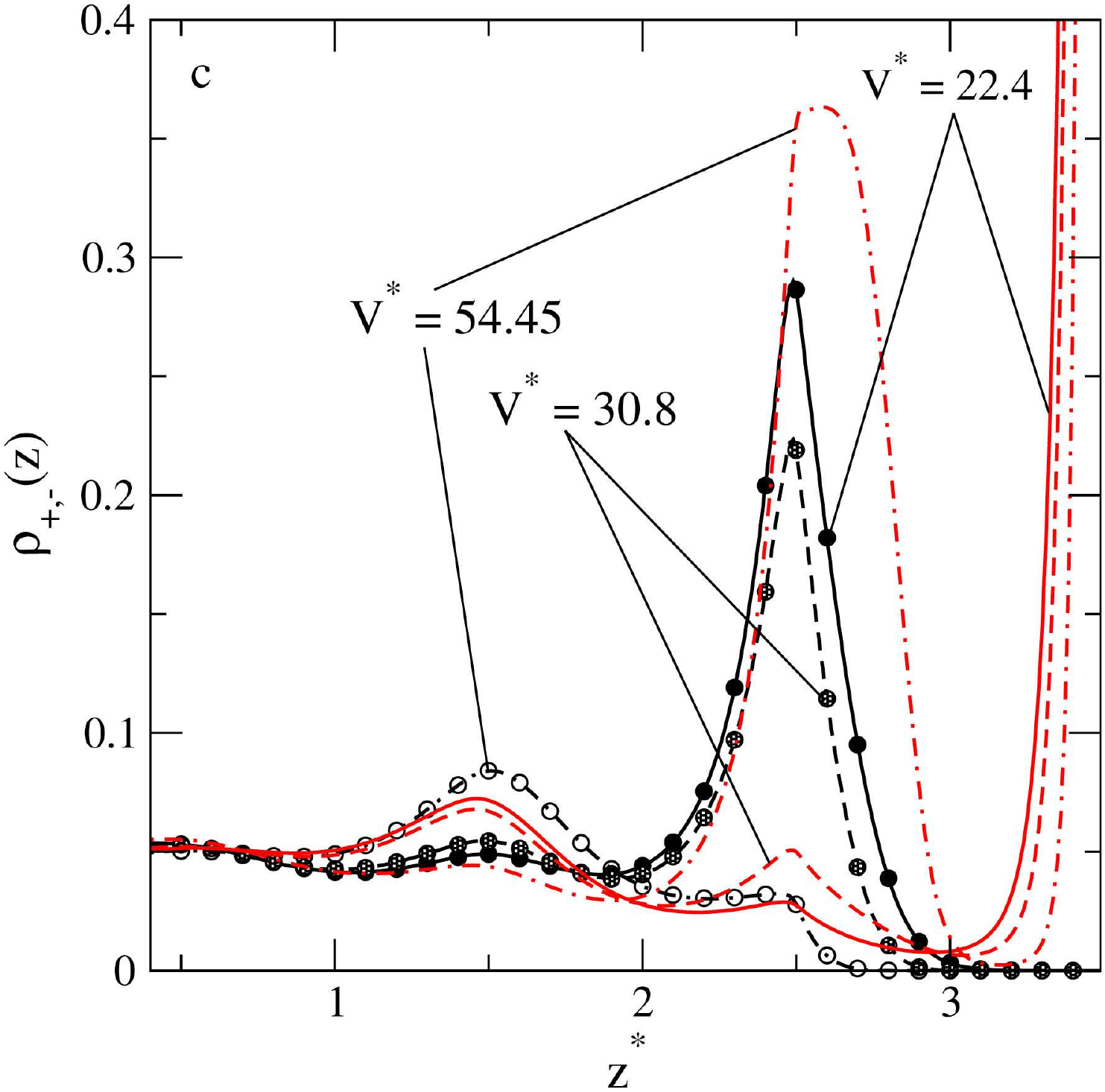}
\end{center}
\vspace{-2mm}
\caption{(Color online) Evolution of the density profiles of each ion species, $\rho_{+,-}(z)$,
of the SPM, with the electrostatic potential applied to the wall, $V^*$.
The system is the same as in figure~\ref{fig1}.\label{fig2}}
\protect
\end{figure}

We begin the discussion by considering the microscopic structure and the resulting thermodynamic
properties. The evolution of the density profile of a hard sphere solvent of the SPM with
an increasing electrostatic potential on the wall is shown in figure~\ref{fig1}.
It can be seen that the contact value of the profile $\rho_\mathrm{hs}(z)$ decreases with
an increasing $V^*$, showing that hard spheres
are expelled from the vicinity of the wall. The density of the second layer at $z^*=2.5$
increases with an increasing $V^*$, reaches a maximum value at $V^*=30.8$ and then slightly
decreases with a further increase of
the electrostatic potential. This proves that
hard sphere solvent particles are again slightly expelled from the second layer at the expense of
a weakly increasing density closer to the pore center. These trends are due to accumulation and
simultaneous separation of ion species close to the charged surface of the pore, figure~\ref{fig2}.

The density of counter-ions substantially increases close to the pore walls while the density of
co-ions decreases at the contact and in the pore walls vicinity with an increasing $V^*$.
However, structural changes also occur in the second layer around $z^*=2.5$. In this layer,
the co-ion density increases while opposite trends are seen for the counter-ions. It seems,
however, that the presence of hard sphere solvent species in the pore center promotes the separation
of ions of the opposite charge close to the wall, thus playing
a role of supporting the external field effects.

\begin{figure}[!t]
\begin{center}
\includegraphics[width=0.425\textwidth,clip]{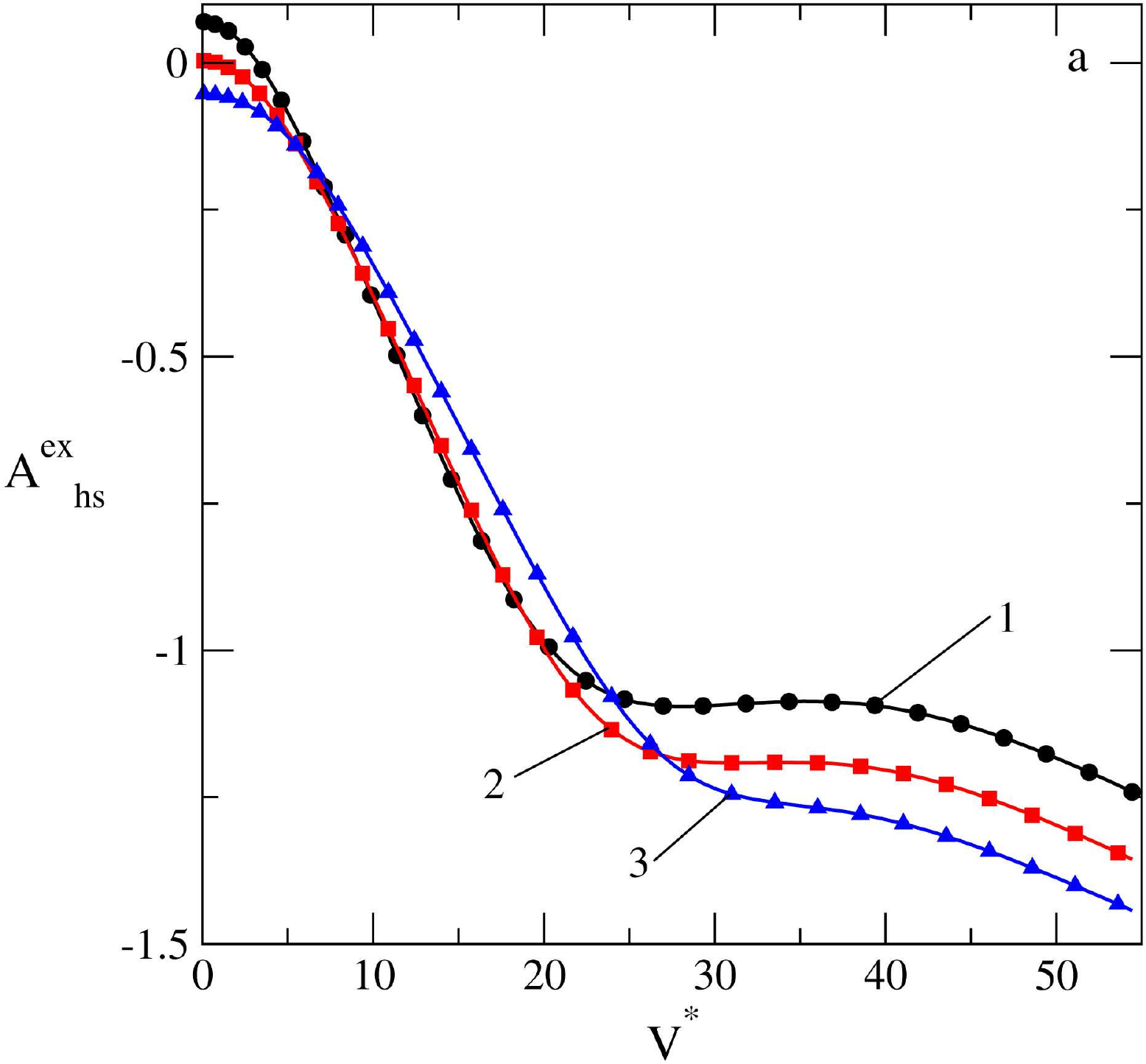}\hspace{3mm}
\includegraphics[width=0.42\textwidth,clip]{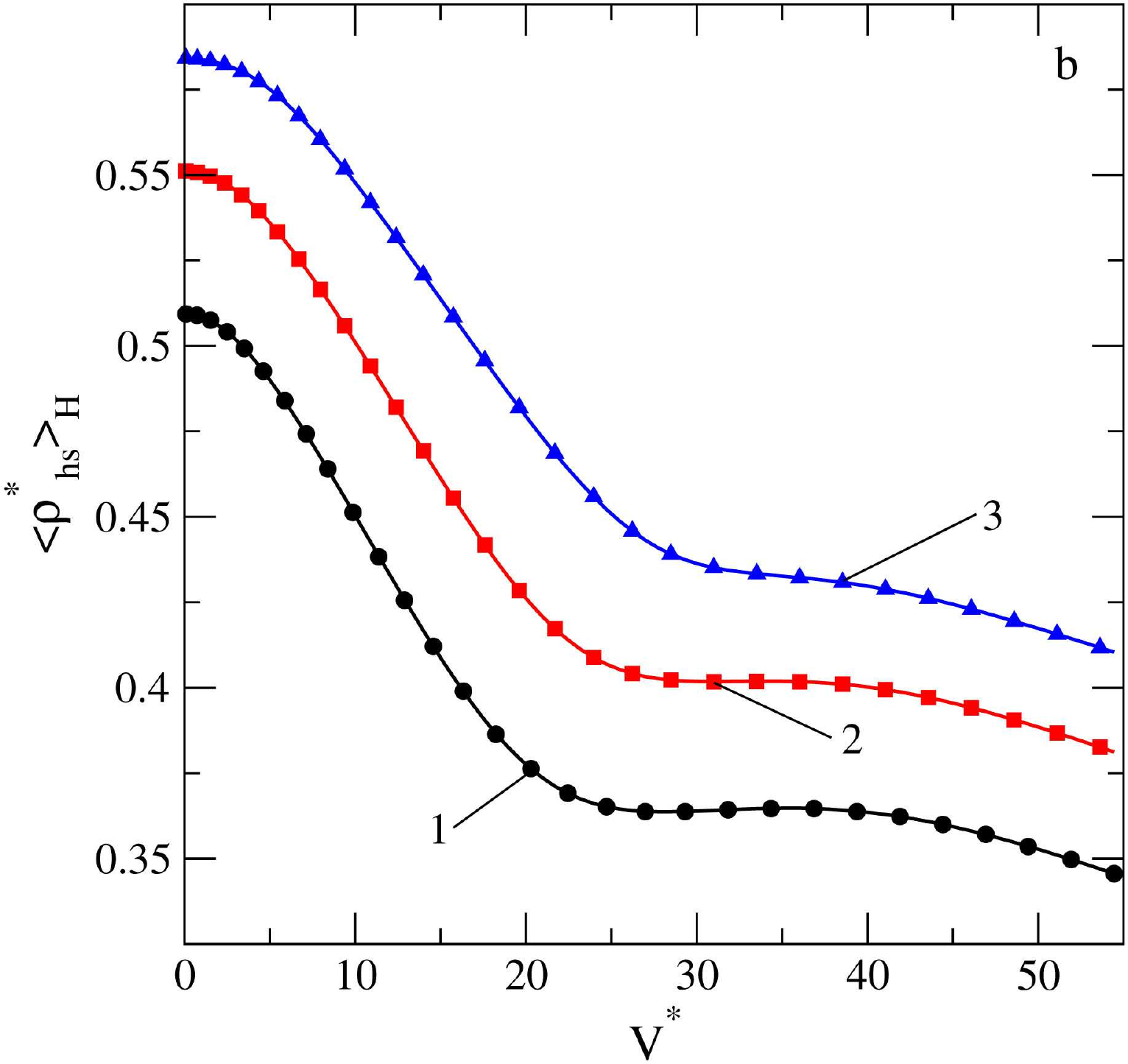}\\[1ex]
\includegraphics[width=0.43\textwidth,clip]{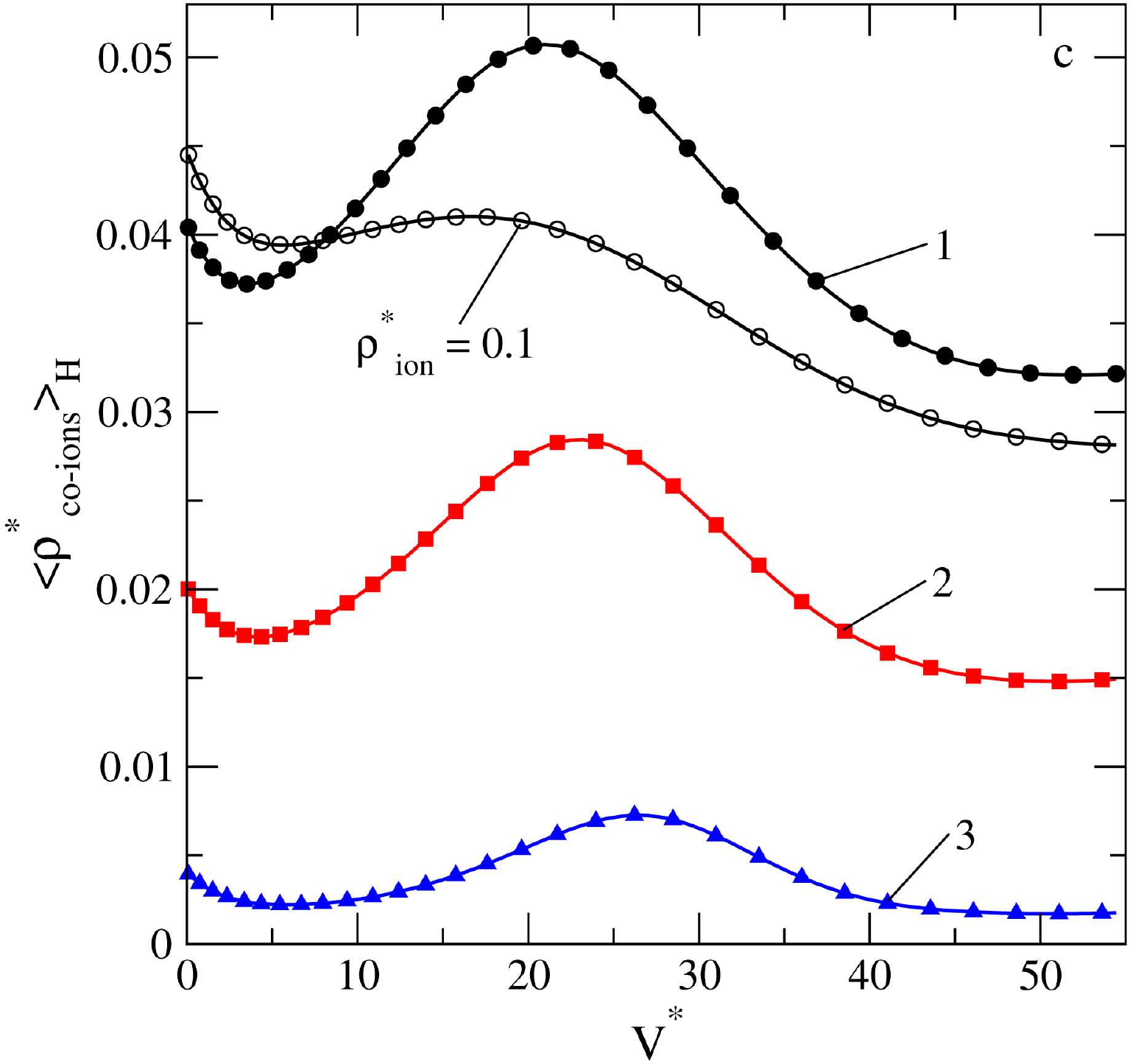}\hspace{3mm}
\includegraphics[width=0.42\textwidth,clip]{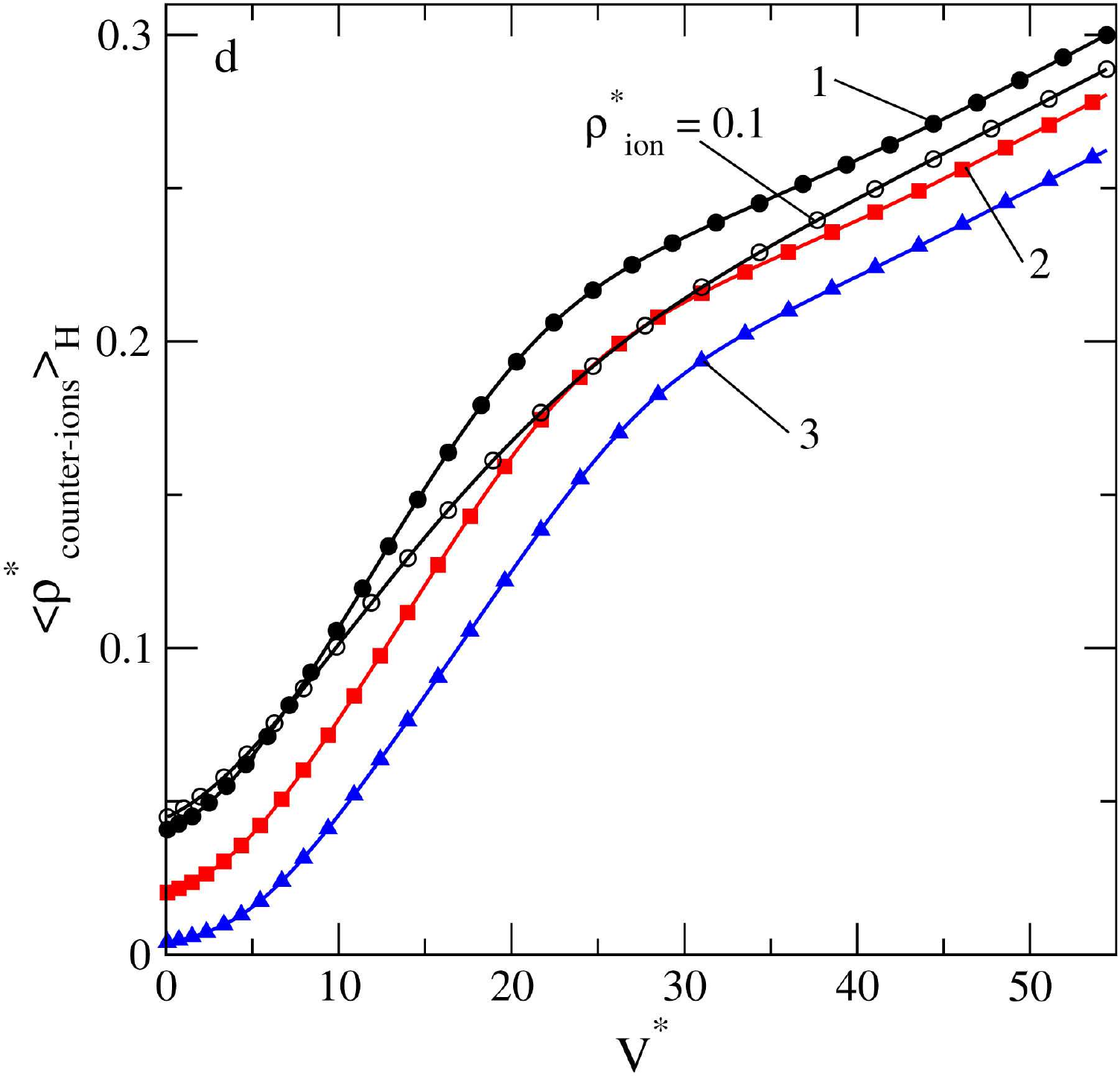}
\end{center}
\vspace{-3mm}
\caption{(Color online) Excess adsorption and average density of hard sphere species of the SPM
in the slit like pore $H^*=8$, panels (a) and (b), respectively.
Average density of co-ions [panel (c)] and of counter-ions [panel (d)] in this pore.
The nomenclature of systems is the following: 1~--- $\rho^*_\mathrm{hs}=0.5$, $\rho^*_\mathrm{ion}=0.1$;
2~--- $\rho^*_\mathrm{hs}=0.55$, $\rho^*_\mathrm{ion}=0.05$; 3~--- $\rho^*_\mathrm{hs}=0.59$, $\rho^*_\mathrm{ion}=0.01$.\label{fig3}}
\protect
\end{figure}

The trends of behavior of the average density of species in the pore with an increasing electrostatic
potential are illustrated in figure~\ref{fig3}. Excess adsorption of the species is defined as common:
\begin{equation}
A^\mathrm{ex}_{\alpha}=\int \rd z \left[\rho_{\alpha}(z)-\rho_{\alpha}\right]
\end{equation}
and the average density of the species is as follows:
\begin{equation}
\langle \rho_{\alpha}\rangle_H = \frac{1}{H} \int \rd z \rho_{\alpha}(z).
\end{equation}
From the panel (a) of this figure we learn that the excess adsorption of a hard sphere solvent
substantially decreases with an increasing electrostatic potential and is negative
in almost entire range of $V^*$.
In all three cases considered,
we kept constant the total density of the bulk solution at $\rho_\mathrm{hs}+\rho_\mathrm{ion}=0.6$ and
changed
its
composition by decreasing the ion density in the systems 1, 2, and 3. The curves behave
differently at low, intermediate and high $V^*$. In
a narrow region of rather small $V^*$
and at high values of $V^*$, the lowest excess adsorption is observed for the system $3$
that has
the
lowest fraction of ions in the bulk phase. The average density of hard sphere species
[panel (b)] decreases with an increasing $V^*$, its dependence on $V^*$ being non-monotonous, however.

This behavior of the excess adsorption and of the average density of solvent species is due to
the changes of the average density (and distribution of ions) in the pore under the effect of
external electric field. In particular, the behavior of the average density of co-ions [panel (c)] with
an increasing $V^*$ having a maximum in the interval between 20 and 25 can be traced back to the
corresponding density profiles showing how the co-ions are expelled from the vicinity of the
wall and how they form a relatively dense second layer. Again, changes of the structure discussed
in terms of the curves in figure~\ref{fig2} are manifest due to the different rate of growth at low and high
$V^*$ of the average density of counter-ions [panel (d) of figure~\ref{fig3}].
Most important, changes of the density of ion species and changes of distribution of solvent
species upon increasing the electrostatic potential cause the changes of the dependence of the charge
in the pore and consequently the changes in the shape of the differential capacitance.

\subsection{Differential capacitance}

\begin{figure}[!t]
\begin{center}
\includegraphics[width=0.45\textwidth,clip]{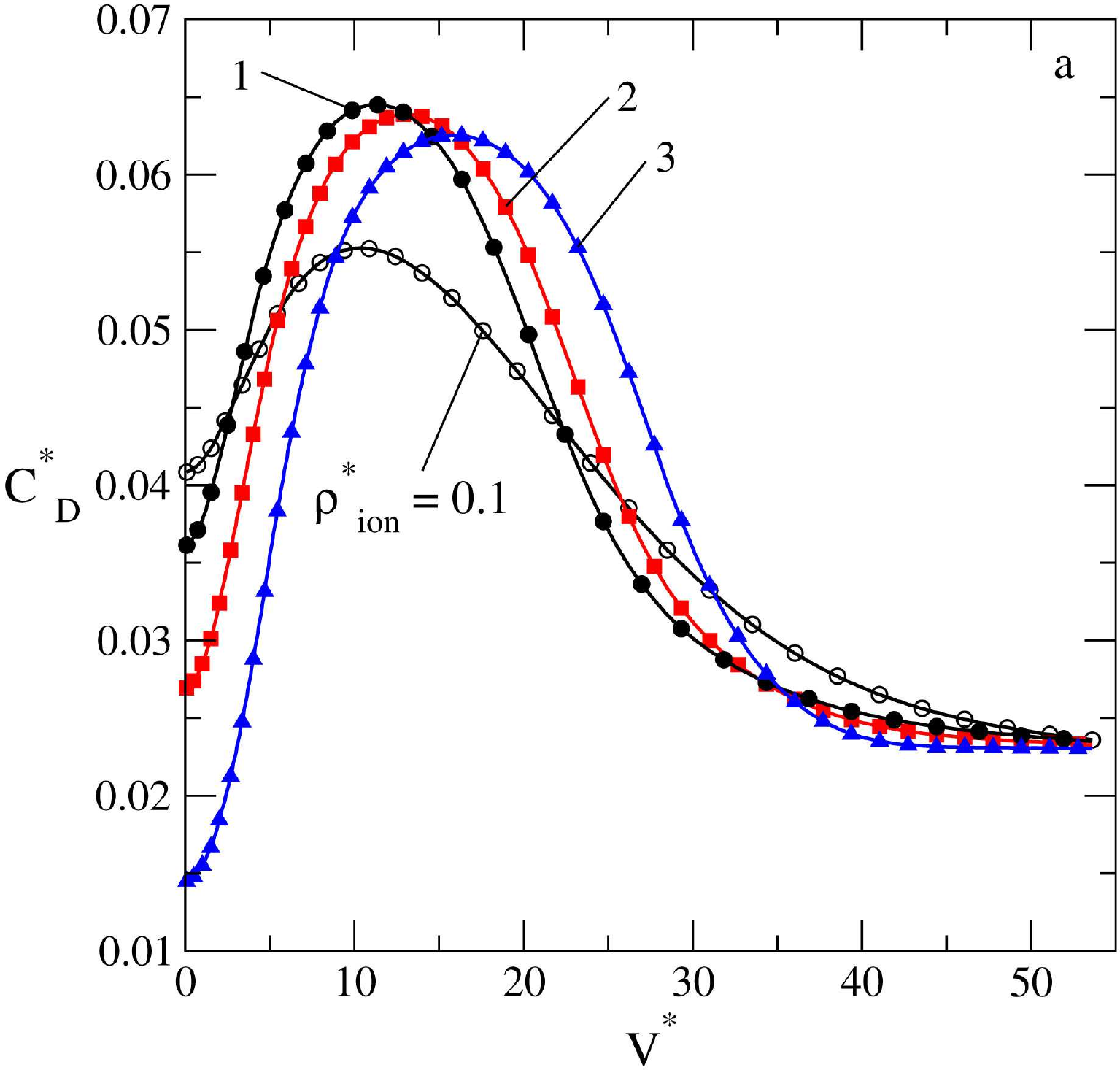}\\[1ex]
\includegraphics[width=0.45\textwidth,clip]{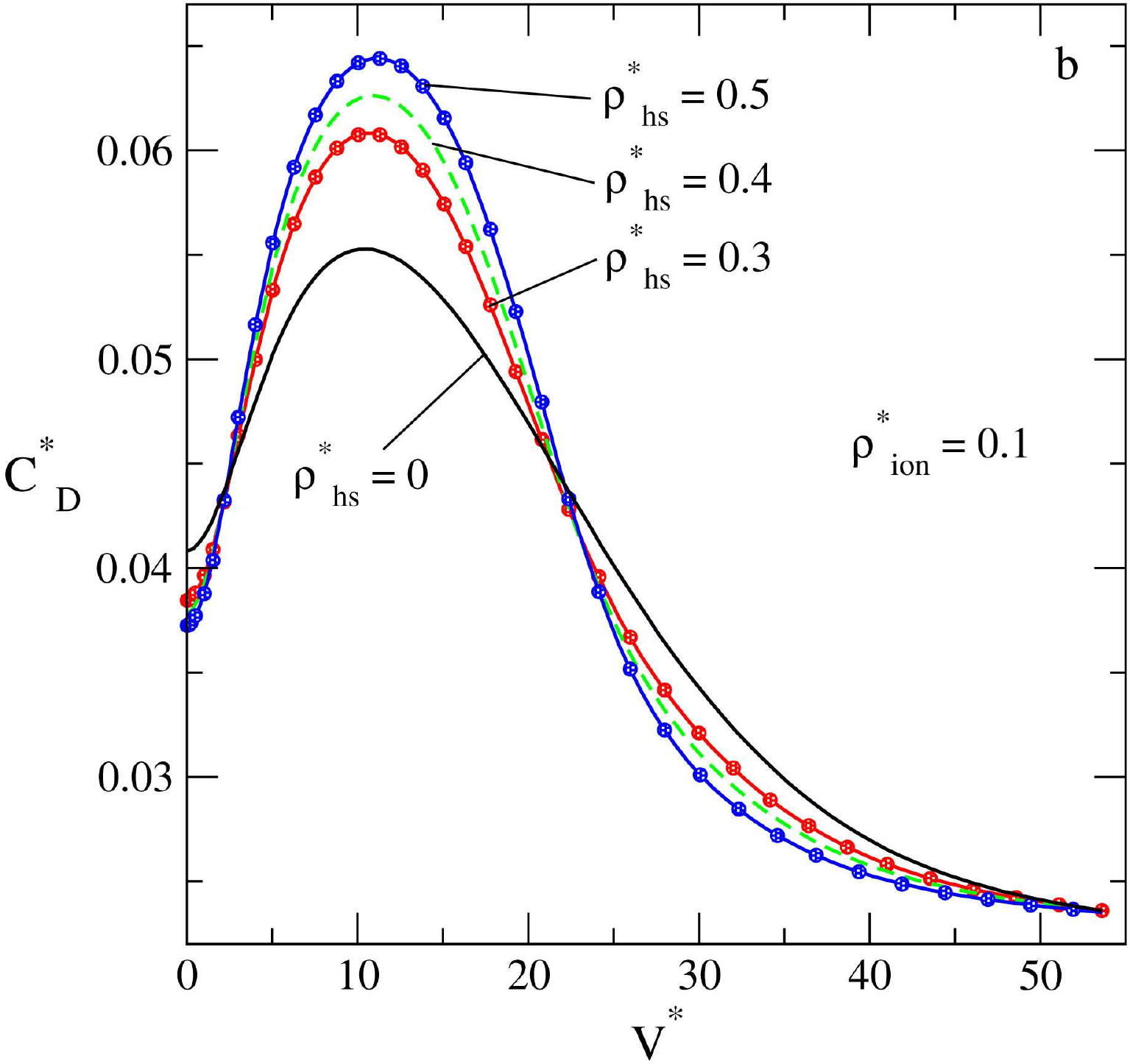}\hspace{2mm}
\includegraphics[width=0.46\textwidth,clip]{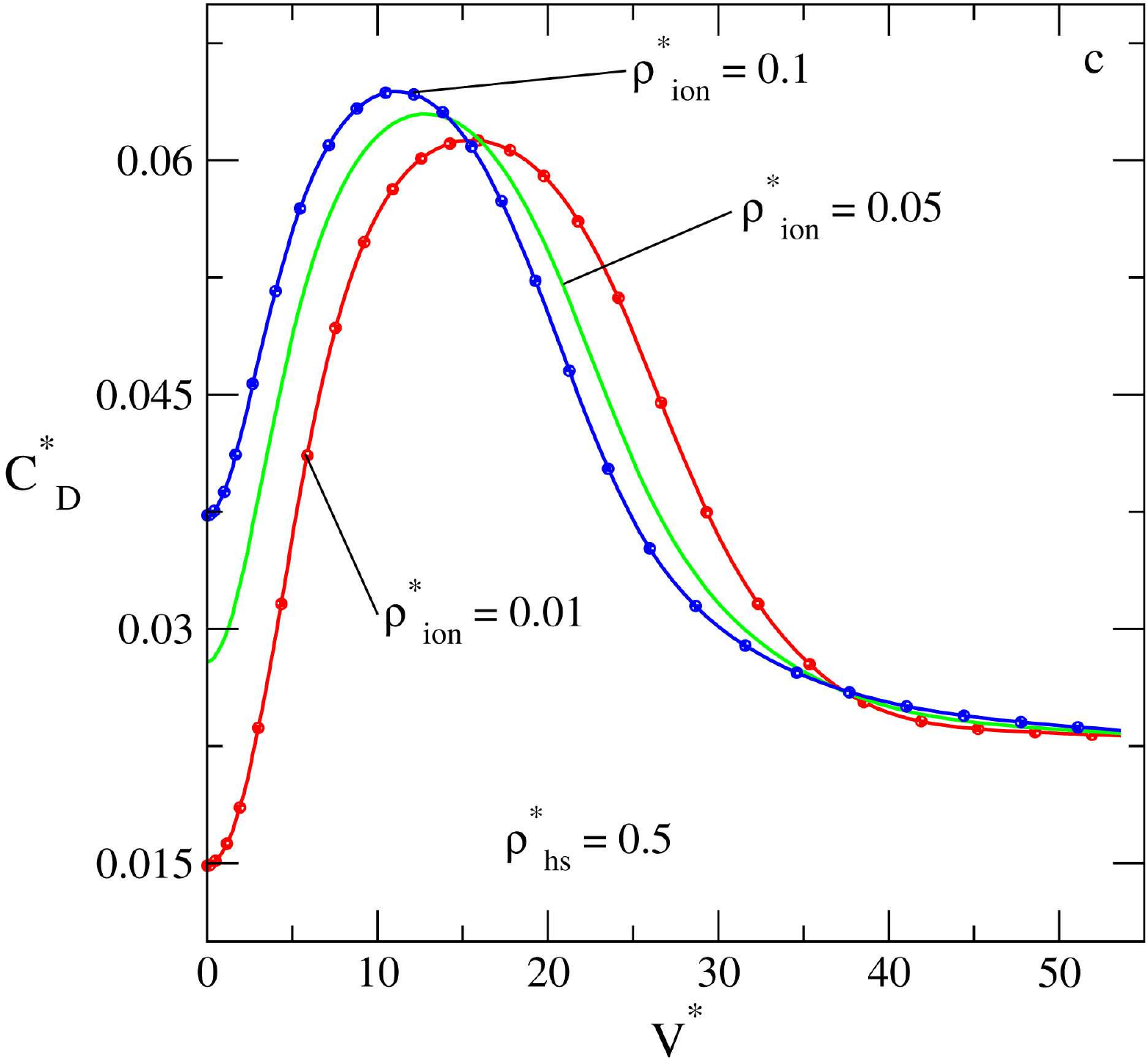}
\end{center}
\vspace{-2mm}
\caption{(Color online) The dependences of the differential capacitance, $C^*_\mathrm{D}$,
on the applied voltage, $V^*$ at a different bulk fluid density and
a different composition shown in each panel. The nomenclature of
systems 1, 2, and 3 [panel (a)] is given in the text.\label{fig4}}
\protect
\end{figure}

The differential capacitance,
\begin{equation}
C_\mathrm{D} = \left(\frac{\partial Q}{\partial V_0}\right)_{H,T,\mu_{\alpha}},
\end{equation}
is obtained by taking a derivative of the charge by electrostatic potential on the wall
and is plotted as a function of electrostatic potential in figure~\ref{fig4}.
In all the cases studied we observe the camel-like shape of the
differential capacitance. Considering the fixed total density as in figure~\ref{fig3}, we see now that
the highest maximum value of the capacitance is reached when the ion fraction is the highest,
namely for the system $1$ compared to $2$ and $3$. However, the value of the maximum is less
sensitive to the ion fraction compared to the trough at a very small $V^{*}$ [panel (a) of figure~\ref{fig4}].
At  a very high $V^*$
the curves for three systems tend to almost equal value. If we compare the system $1$ and its
PM counterpart at the same ion density ($\rho^*_\mathrm{ion}=0.1$), then it appears that the differential
capacitance curves behave qualitatively similarly. However, in the SPM case, the $C^*_\mathrm{D}$ maximum
is much higher compared to PM. Thus, it seems that the presence of solvent species enhances
the separation of ions of the opposite charge by ``putting'' them slightly closer to the pore walls,
where the electric field makes its job. In order to obtain higher values of the differential
capacitance at maximum, one can either take a denser solvent (at a fixed ion density) like it is shown in
the panel (b) of figure~\ref{fig4} or may increase the ion fraction at a fixed solvent density, like in the panel (c)
of figure~\ref{fig4}. To summarize, the presence of solvent species in the SPM
permits to alter the values of differential capacitance
in different regions of $V^*$, in comparison to PM. However, the overall shape remains
qualitatively similar unless the ion fraction becomes high (in real  systems one needs in fact
to take into account the solubility limit).

\begin{figure}[!t]
\begin{center}
\includegraphics[width=0.45\textwidth,clip]{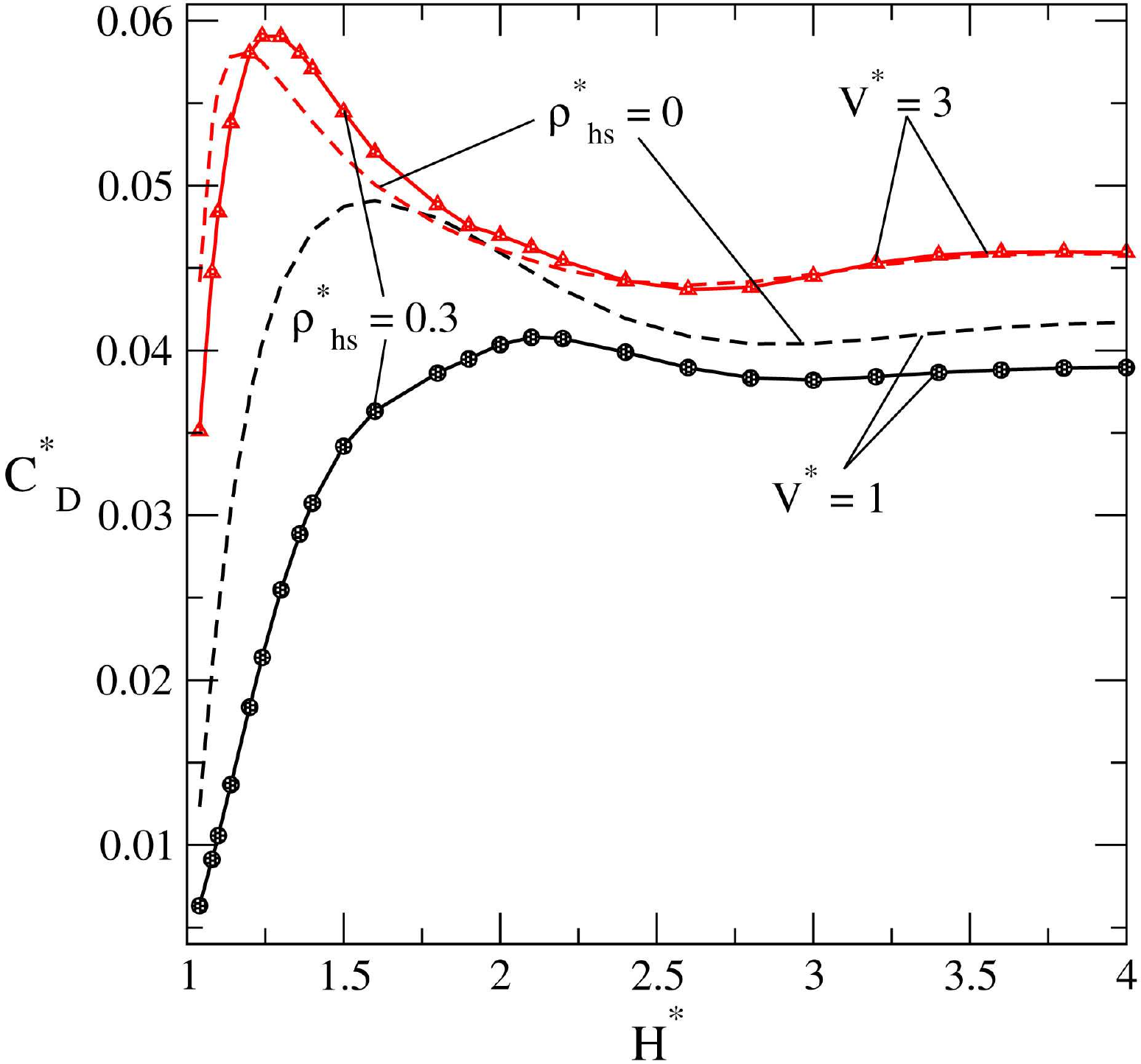}\\[1ex]
\includegraphics[width=0.45\textwidth,clip]{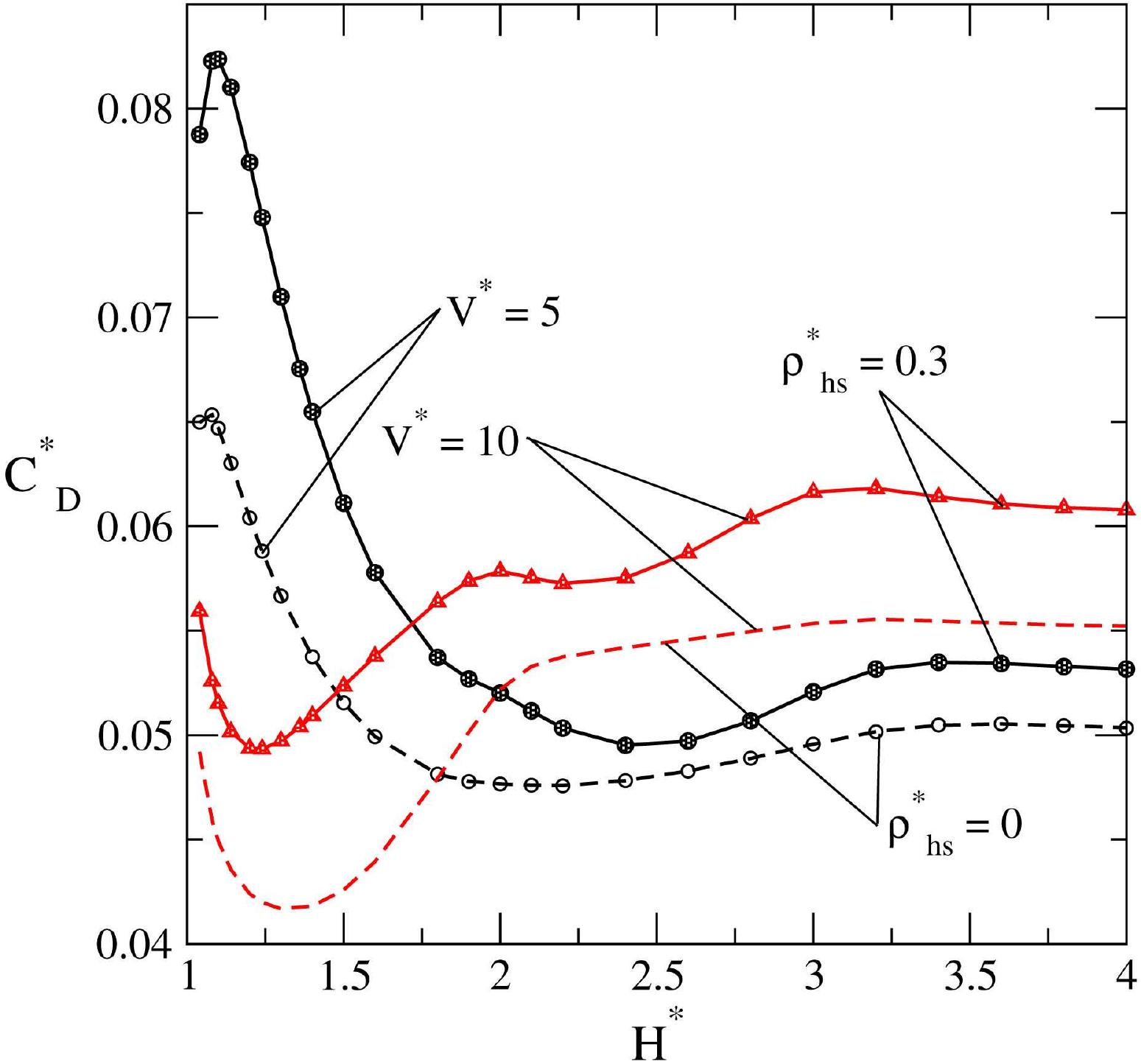}\hspace{3mm}
\includegraphics[width=0.45\textwidth,clip]{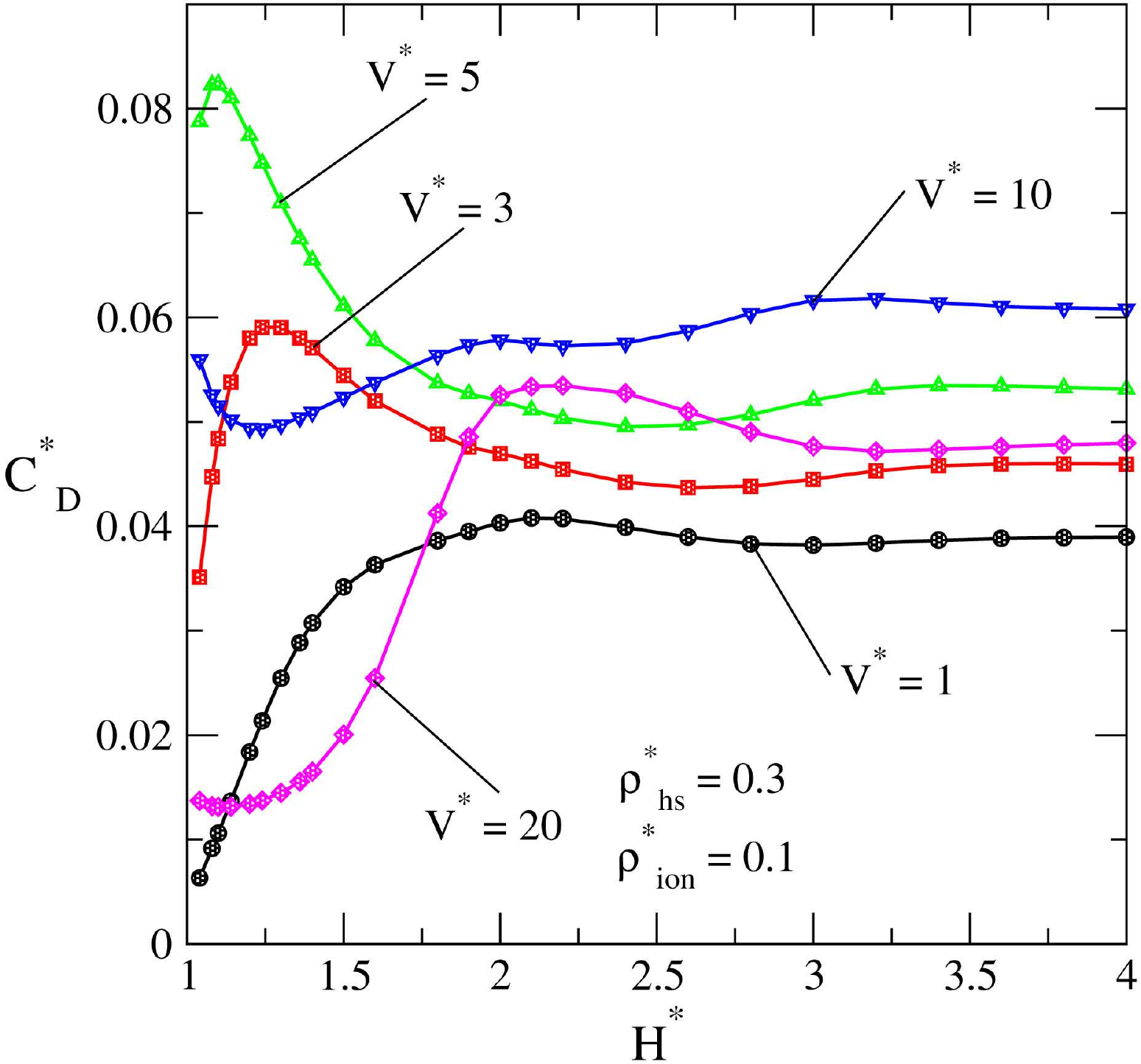}
\end{center}
\vspace{-2mm}
\caption{(Color online) The dependences of the differential capacitance, $C^*_\mathrm{D}$, on the pore width, $H^*$
at a different voltage, $V^*$. Panels (a) and (b) show a comparison of the
results for the SPM and PM at the same conditions.
Panel (c) contains the results for $C^*_\mathrm{D}(H^*)$ solely for the SPM at a
different fixed voltage $V^*$.\label{fig5}}
\protect
\end{figure}

The final issue we would like to discuss
is the dependence of the differential capacitance on the pore width. This problem
for a restricted primitive model of electrolyte solutions confined in slit-like pores
was quite comprehensively discussed  in the recent work from this laboratory~\cite{pizio1}.
In the panel (a) of figure~\ref{fig5} we compare the SPM and PM curves for $C^*_\mathrm{D}(H)$ at rather low values
of the electrostatic potential, namely at $V^*=1$ and $V^*=3$. The curves for two models are of similar shape.  However, the solvent affects the values
for $C^*_\mathrm{D}$, especially in narrow pores. The first maximum of $C^*_\mathrm{D}$ can be either supressed
(at $V^*=1$) or enhanced (at $V^*=3$) due to the solvent presence [figure~\ref{fig5}~(a)].
The curves for SPM and PM eventually
tend to zero if $H^*$ tends to its minimum value.

At  higher values of $V^*$, $V^*=5$ and $V^*=10$ [panel (b) of figure~\ref{fig5}],
the qualitative features of the shape of functions in question are again similar for SPM and PM.
Nevertheless, in the SPM case, we observe more pronounced oscillations of the differential
capacitance on the pore width. In other words, the phase of overlap of the density profiles of ions
formed at each wall (discussed in detail in ~\cite{pizio1}) is altered, due to the presence of
solvent species. In close similarity to the PM system, the shape of the dependence of $C^*_\mathrm{D}(H^*)$
in the present SPM case alters
depending on the value of the electrostatic
potential $V^*$. The differential capacitance can either grow or drop in the region
of very narrow pores  depending
on the choice of the voltage. Still, the oscillatory
behavior (showing well pronounced and less pronounced maxima  and several troughs) is observed for the confined
SPM.

It is interesting to mention that Oleksy and Hansen observed the
oscillatory curve for the solvation force between charged plates with the SPM-like solution
in between. However, their calculations were performed under the condition of a constant charge on the plates
rather than at a constant potential carried out in the present study.
It seems that to establish the relation between the oscillatory curve for the differential
capacitance and the dependence of the solvation force on the charged plates separation
is of utmost importance in future research.
In addition, we would like to emphasize that the model of this study permits several extensions.
One of the promising extensions is the possibility
to improve the model by introducing the concepts of chemical association in order to deal with the adsorption of
either chain molecules or the network-forming solvent. Theoretical background is rather
straightforward to be developed along the lines presented in e.g.,~\cite{bryk2}.

\section*{Acknowledgements}
O.P. is grateful to David Vazquez for technical assistance at the Institute of Chemistry
of the UNAM.

\ukrainianpart
\title{Примітивна модель розчинника подвійного електричного шару в щілиноподібній порі: мікроскопічна структура,
адсорбція та електроємність з використанням методу функціоналу густини}

\author[O. Пізіо, С. Соколовскі]{O. Пізіо\refaddr{label1}, С. Соколовскі\refaddr{label2}}
\addresses{
\addr{label1} Інститут хімії, Національний автономний університет м. Мехіко, Мехіко, Мексика
\addr{label2} Відділ моделювання фізико-хімічних процесів, Університет Марії Кюрі-Склодовської, Люблін, Польща}

\makeukrtitle

\begin{abstract}
Ми досліджуємо електричний подвійний шар, сформований між двома зарядженими стінками щілиноподібної пори, і примітивну модель розчинника
для розчину електроліту. Недавно розвинута версія методу зваженого функціоналу густини для електростатичної міжчастинкової взаємодії
застосовується до вивчення профілів густини, адсорбції і селективності адсорбованих іонів і компонентів розчинника. Ми
звертаємо основну нашу увагу, проте, на залежність диференційної електроємності від прикладеної напруги на електродах і  в порі.
Ми обговорюємо властивості моделі по відношенню до поведінки примітивної моделі, а саме, у відсутності твердокулькового розчинника.
Ми спостерегли, що диференційна електроємність примітивної моделі розчинника при прикладеному електростатичному потенціалі
має ``двогорбову'' форму, незважаючи на високу концентрацію іонів. Крім того, знайдено, осцилюючу залежність диференційної ємності
примітивної моделі розчинника від ширини пори, що є  дуже подібно до примітивної моделі.

\keywords примітивна модель розчинника, функціонал густини, розчини електролітів, адсорбція, диференційна електроємність

\end{abstract}
\end{document}